\documentclass{article}

\usepackage{arxiv}

\usepackage{bm}
\usepackage{subcaption}
\usepackage{amsmath}
\usepackage{amsfonts}
\usepackage{listings, newtxtt}
\usepackage{graphicx}
\usepackage{xcolor}
\usepackage{pgffor} 

\usepackage{booktabs}
\usepackage{tabularx}
\usepackage{placeins} 
\usepackage[]{units}

\newcommand{\orcidicon}{\includegraphics[width=0.32cm]{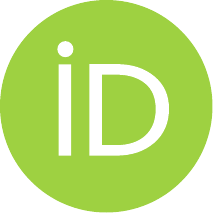}}

\foreach \x in {A, ..., Z}{%
\expandafter\xdef\csname orcid\x\endcsname{\noexpand\href{https://orcid.org/\csname orcidauthor\x\endcsname}{\noexpand\orcidicon}}
}

\title{pyCFS-data: Data Processing Framework in Python for openCFS}

\author{ Andreas Wurzinger\orcidA{}\textsuperscript{1,2,}\thanks{These authors contributed equally to this work and share the first authorship.} \\
	\textsuperscript{1}Aero- and Vibroacoustics Group, IGTE\\
	Graz University of Technology\\
	Inffeldgasse 18, 8010 Graz, Austria \\[0.6ex]
	\textsuperscript{2}Institute of Mechanics\\
	Graz University of Technology\\
	Kopernikusgasse 24, 8010 Graz, Austria \\[0.6ex]
	\texttt{andreas.wurzinger@tugraz.at} \\
	\And
	Patrick Heidegger\orcidB{}\footnotemark[1] \\
	Aero- and Vibroacoustics Group, IGTE\\
	Graz University of Technology\\
	Inffeldgasse 18, 8010 Graz, Austria \\
	\texttt{patrick.heidegger@tugraz.at} \\
	\And
	Stefan Schoder\orcidC{} \\
	Aero- and Vibroacoustics Group, IGTE\\
	Graz University of Technology\\
	Inffeldgasse 18, 8010 Graz, Austria \\
	\texttt{stefan.schoder@tugraz.at} \\
}

\renewcommand{\headeright}{Technical Report}
\renewcommand{\undertitle}{Technical Report}
\renewcommand{\shorttitle}{pyCFS-data}

\usepackage[backend=biber, style=numeric, sorting=nyt, maxbibnames=99]{biblatex}
\usepackage[
    bookmarksopen = true, 
    colorlinks=true,      
    linkcolor=teal,       
    citecolor=teal,       
    filecolor=teal,    
    urlcolor=teal,      
    plainpages = false,
    hypertexnames = false,
    ]{hyperref}
\hypersetup{
pdftitle={pyCFS-data: Data Processing Framework in Python for openCFS},
pdfsubject={Technical report},
pdfauthor={Wurzinger, Heidegger, Schoder},
pdfkeywords={Data processing framework, Python, Open source software, Simulation data processing, Data manipulation, openCFS},
}

\usepackage{cleveref}

\begin{document}
\maketitle

\begin{abstract}
	openCFS is an open-source finite element software for multiphysics simulations, e.g., aero- and vibroacoustics.
	This paper presents pyCFS-data, a Python-based data processing framework for pre- and post-processing of data that is generated by, or used for, openCFS.
	The framework reads and writes the openCFS HDF5 file format into NumPy-based data structures for meshes and results, and converts data from other formats, e.g., EnSight Gold and CGNS files of CFD simulations, Ansys and Exodus files, and universal files of laser vibrometer measurements.
	It provides a range of operators, e.g., to interpolate data between meshes and between nodes and elements, compute derivatives in time and space, perform mesh-transformation operations, and construct synthetic sound sources.
	Operations on large data, such as the computation of aeroacoustic sources, are supported by a pipeline structure that processes data sets chunk by chunk within a specified memory limit.
	The pipeline is checked for metadata consistency before any bulk data is read, with multi-level dry-run options available. This report describes pyCFS~0.3.0, which added the pipeline module, spatial, convective, and substantial derivatives, conservative interpolation, and time averages.
\end{abstract}

\keywords{Data processing framework \and Python \and Open source software \and Simulation data processing \and Data manipulation \and openCFS}

\begin{table*}[htbp!]
	\centering
	\caption*{Code metadata.}
	\small
	\begin{tabular}{@{}lp{0.38\textwidth}p{0.5\textwidth}@{}}
		\toprule
		Nr. & Code metadata description                                         & Metadata                                                                                                                                                    \\
		\midrule
		C1  & Current code version                                              & 0.3.0                                                                                                                                                       \\
		C2  & Permanent link to code/repository used for this code version      & \url{https://gitlab.com/openCFS/pycfs}                                                                                                                      \\
		C3  & Legal code license                                                & MIT License                                                                                                                                                 \\
		C4  & Code versioning system used                                       & git                                                                                                                                                         \\
		C5  & Software code languages, tools and services used                  & Python ($\geq$\,3.10)                                                                                                                                       \\
		C6  & Compilation requirements, operating environments and dependencies & CI-tested: Linux and Windows;\newline NumPy, SciPy, Numba, h5py, netCDF4, and further packages; \newline optional: VTK (EnSight), meshio, Ansys DPF, PyDMD; \\
		C7  & Link to developer documentation/manual                            & \url{https://opencfs.gitlab.io/pycfs/}                                                                                                                      \\
		C8  & Support email for questions                                       & Issue tracker:\newline \url{https://gitlab.com/openCFS/pycfs/-/issues}                                                                                      \\
		\bottomrule
	\end{tabular}
\end{table*}

%
\FloatBarrier
\section{Motivation and significance}
\label{sec:Intro}

Within this contribution, we concentrate on the \textit{pyCFS}~\cite{Pycfs2026} module pyCFS-data, a data processing framework for the open-source finite-element (FE) simulation code openCFS~\cite{OpenCFS,mayrhoferOpenCFSOpensourceFiniteelement2026}.
Typical tasks of pyCFS-data are the preparation (pre-processing) of input data for openCFS FE simulations and evaluation (post-processing) of openCFS result files. For this purpose, pyCFS-data provides a range of functionality to facilitate the manipulation of meshes and data, reading/converting from alternative file formats, and chunk-processing of large data sets.
In contrast to other general-purpose tools, pyCFS-data reads and writes the openCFS file format with its regions, result types, and MultiSteps.
ParaView visualizes openCFS files with the openCFS reader plugin, cf.~\cref{sec:architecture}.
Other file formats can be imported, e.g., via meshio and VTK, which are optional dependencies of pyCFS-data.
pyCFS-data consists of four main components, which are organized in same-titled submodules,
\begin{itemize}
	\item \texttt{io},
	\item \texttt{operators},
	\item \texttt{extras}, and
	\item \texttt{pipeline}.
\end{itemize}
The \texttt{io} submodule focuses on reading, writing, and accessing data in HDF5 file format as used in openCFS. The \texttt{operators} submodule contains functionality for data pre- and post-processing based on data structures defined in \texttt{io} submodule. The \texttt{extras} submodule contains methods to interact with data stored in different file formats.
The \texttt{pipeline} submodule chains \texttt{operators} into filters that are sequentially applied in a pipeline, which processes large data sets chunk by chunk.
The modules are complemented by utility functions in a top-level file \texttt{util.py}, e.g., for the mesh connectivity, element volumes and quality, and the trilateration of measurement positions.

This report corresponds to \textit{pyCFS}~0.3.0, which introduced the pipeline submodule alongside spatial, convective, and substantial derivatives, conservative cell-centroid interpolation, and time averages. The version also covers performance improvements, extends the continuous integration (CI) test suite, and revised code parts like the node-to-cell and nearest neighbor interpolators, and the VTK-based EnSight reader.

The focus of this document is on the software architecture and the methods behind the operators, complementing the \textit{pyCFS} user documentation~\cite{PyCFSDocumentation}.
\Cref{sec:software} describes the software architecture in \cref{sec:architecture}, the data structures and I/O in \cref{sec:io}, the available operators in \cref{sec:operators}, the data processing pipeline in \cref{sec:pipeline}, and the interfaces to other data formats in \cref{sec:extras}. \Cref{sec:example} illustrates the pipeline with the computation of an aeroacoustic source term, \cref{sec:impact} summarizes applications of pyCFS-data, and \cref{sec:conclusions} concludes the report.

\paragraph{Installation}
The module can be easily installed, including all dependencies, via PyPI\footnote{https://pypi.org/project/pyCFS/}
\begin{lstlisting}[language=Bash]
pip install pycfs
\end{lstlisting}

Detailed instructions for the installation inside a virtual environment and for the developer setup can be found in the user documentation~\cite{PyCFSDocumentation}. Optional dependencies are installed with the extras: \texttt{vtk} for reading EnSight files, \texttt{meshio}, \texttt{dmd} for dynamic mode decomposition with PyDMD, and \texttt{ansys} for the Ansys DPF package. All available dependencies are installed via \texttt{pycfs[all]}.
Note that some features, e.g., the conservative interpolation and the spatial derivatives, use Numba-compiled kernels, which are cached on disk. In read-only installations, set the environment variable \texttt{NUMBA\_CACHE\_DIR} to select a writable cache directory.


\section{Software description}
\label{sec:software}

\subsection{Software architecture}
\label{sec:architecture}

\begin{figure}[htbp!]
	\centering
	\includegraphics[width=0.9\textwidth]{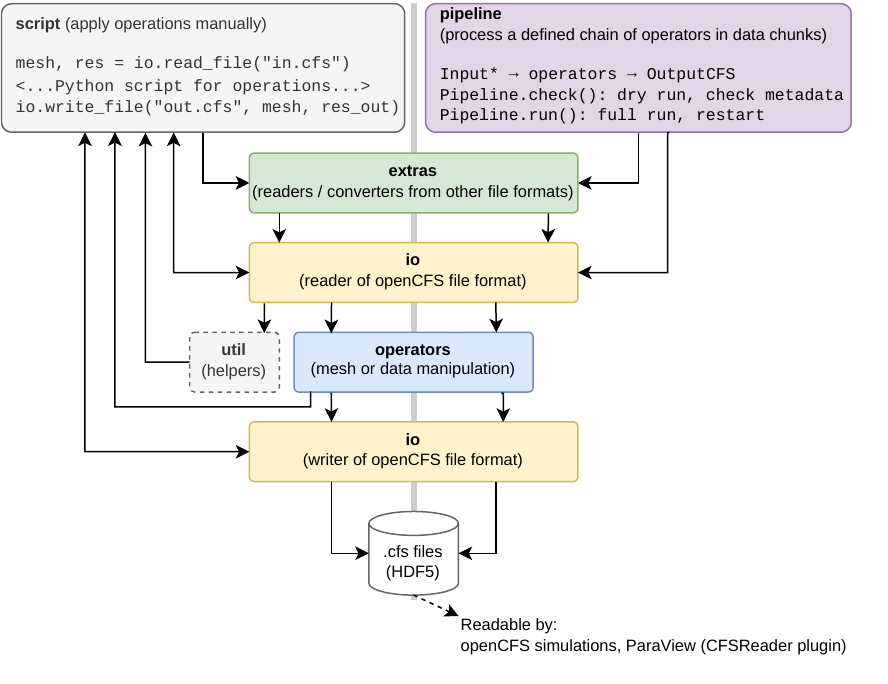}
	\caption{Submodules of pyCFS-data and the two ways to use them: as API in a \texttt{script}, or as a \texttt{pipeline}.}
	\label{fig:architecture}
\end{figure}

The four submodules of pyCFS-data build on each other, as illustrated in~\cref{fig:architecture}.
The data structures of \texttt{io} hold the mesh and the result data of all other submodules. The \texttt{extras} submodule converts other file formats into them. The operators submodule covers functions on the \texttt{io} data structures.
Generally, the pyCFS-data routines can be used in two ways: in a custom Python \texttt{script} directly tapping into the application programming interface (API), or using the \texttt{pipeline} submodule.
A \texttt{script} offers most flexibility, using any Python code to process the data. Reader/writer operations, data manipulations, and application of \texttt{operators} are done manually in this case.
In contrast, a \texttt{pipeline} provides functionality to apply a chain of operators sequentially to the data set, at a high abstraction level. It is designed to process large data sets, where simultaneous processing of all steps is not possible. Here, data processing is done in chunks of (time) steps, which are cut to stay within a specified memory limit.
The output \texttt{.cfs} files can be read by openCFS and visualized in ParaView with the openCFS ParaView reader plugin \texttt{CFSReader}, which is shipped with ParaView since version 5.12 and only needs to be loaded in the plugin manager.


\paragraph{Testing}
The pyCFS-data part is covered by more than 2100 test cases using pytest, running in continuous integration (CI) on Linux for Python 3.10 to 3.13 and on Windows, together with flake8 and mypy checks and a coverage report.
The tests cover all four submodules.
The \texttt{io} tests check the data structures and round trips of writing and reading meshes and results.
Operators are tested against analytic solutions and/or compared with the openCFS testsuite.
The readers of the \texttt{extras} submodule are tested with small files of each format.
The \texttt{pipeline} tests compare the chunked results with those of the operator functions, and cover the dry runs, the memory estimate, and interrupted and restarted runs.

\FloatBarrier
\subsection{Data structures and file I/O (\texttt{io})}
\label{sec:io}

\begin{figure}[htb!]
	\centering
	\begin{subfigure}{\textwidth}
		\centering
		\includegraphics[width=\textwidth]{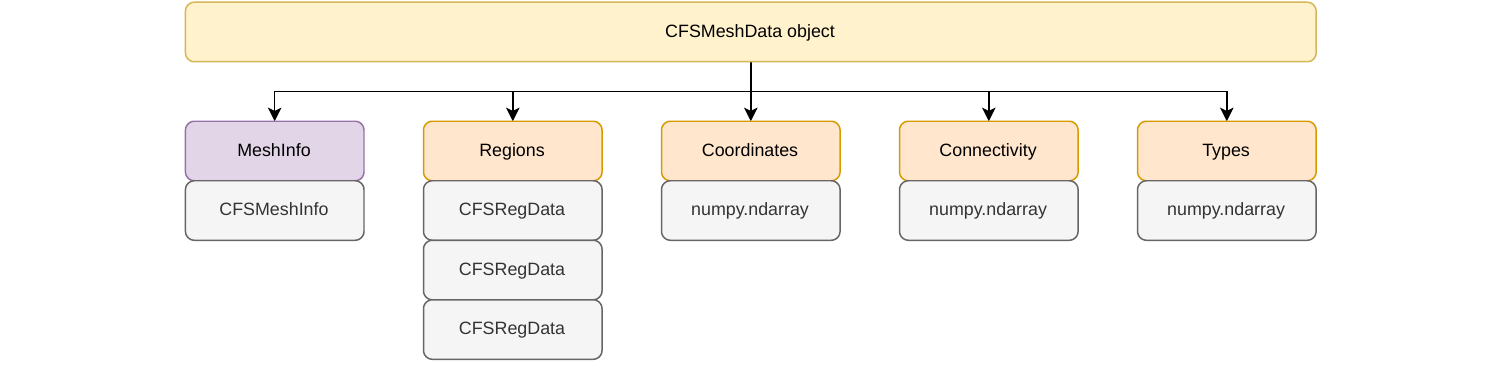}
		\caption{\texttt{CFSMeshData}}
		\label{fig:data_structures_CFSMeshData}
	\end{subfigure}\\[1ex]
	\begin{subfigure}{\textwidth}
		\centering
		\includegraphics[width=\textwidth]{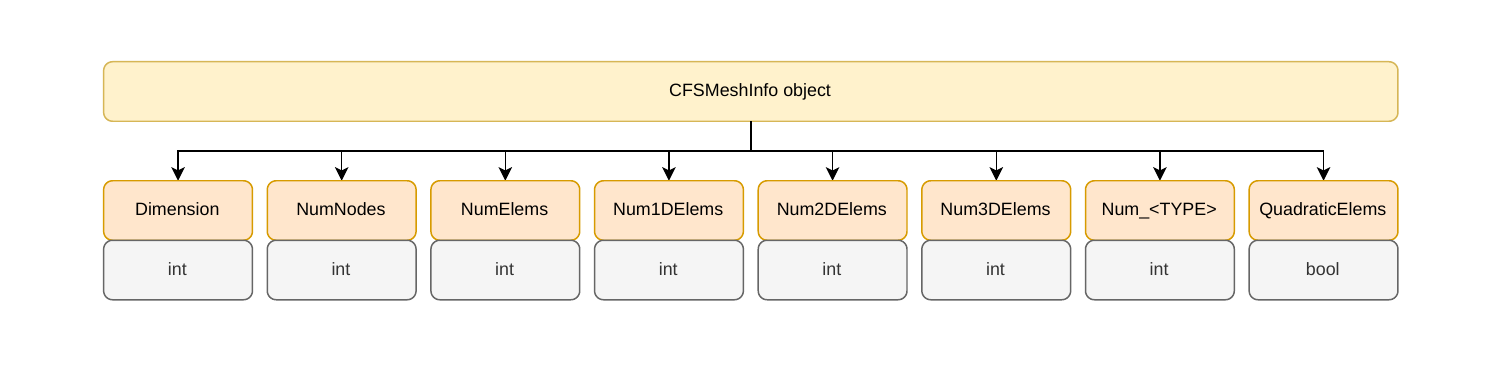}
		\caption{\texttt{CFSMeshInfo}}
		\label{fig:data_structures_CFSMeshInfo}
	\end{subfigure}\\[1ex]
	\begin{subfigure}{\textwidth}
		\centering
		\includegraphics[width=\textwidth]{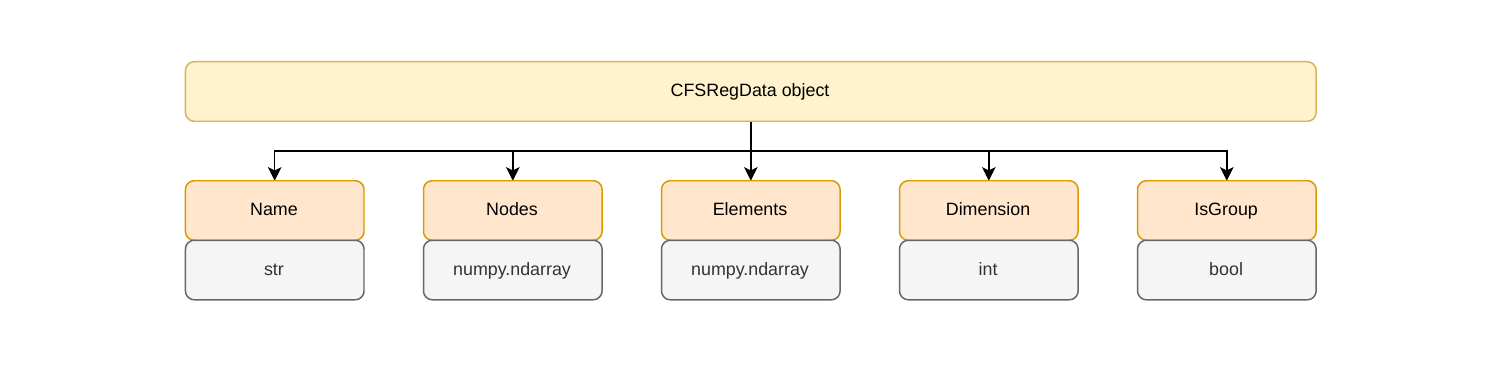}
		\caption{\texttt{CFSRegData}}
		\label{fig:data_structures_CFSRegData}
	\end{subfigure}
	\caption{Mesh data structures. Attributes are highlighted in orange, properties in purple, properties with a setter in an orange-to-purple gradient, and respective data types in grey.}
	\label{fig:data_structures_mesh}
\end{figure}

\begin{figure}[htb!]
	\centering
	\begin{subfigure}{\textwidth}
		\centering
		\includegraphics[width=\textwidth]{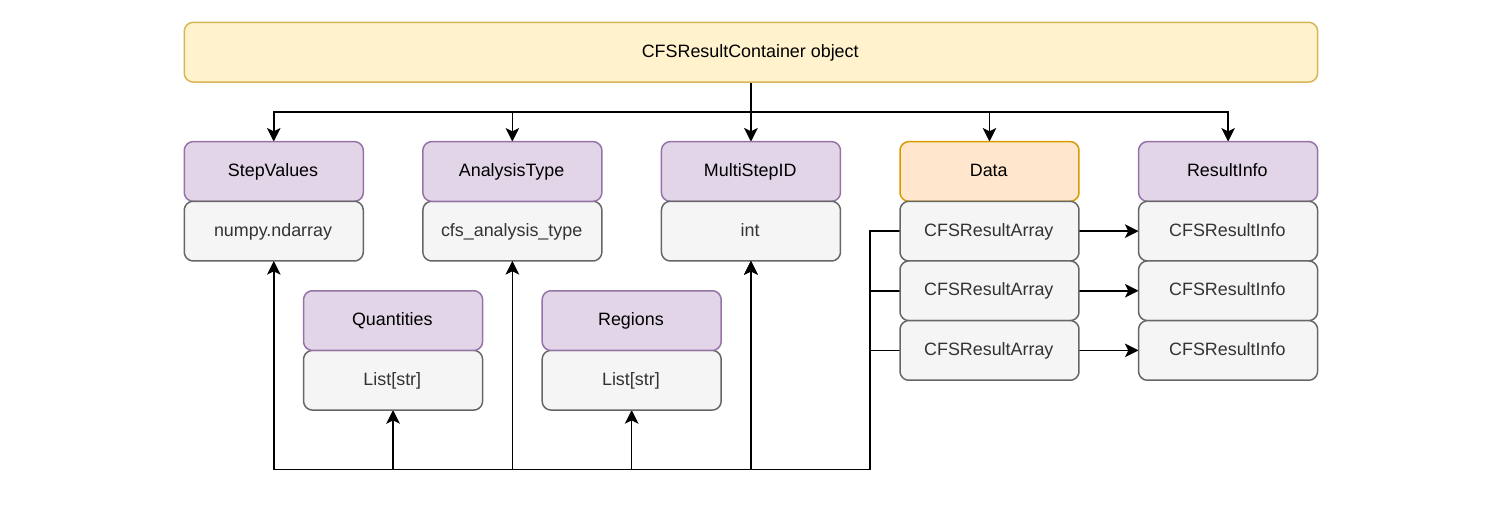}
		\caption{\texttt{CFSResultContainer}}
		\label{fig:data_structures_CFSResultContainer}
	\end{subfigure}\\[1ex]
	\begin{subfigure}{\textwidth}
		\centering
		\includegraphics[width=\textwidth]{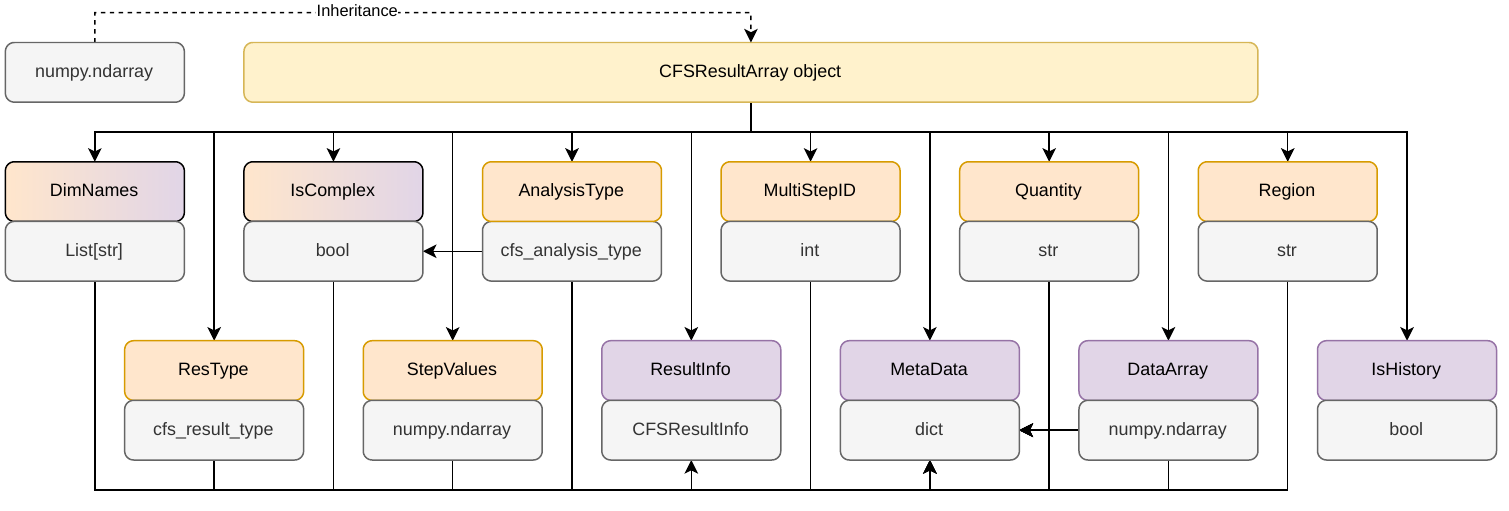}
		\caption{\texttt{CFSResultArray}}
		\label{fig:data_structures_CFSResultArray}
	\end{subfigure}\\[1ex]
	\begin{subfigure}{\textwidth}
		\centering
		\includegraphics[width=\textwidth]{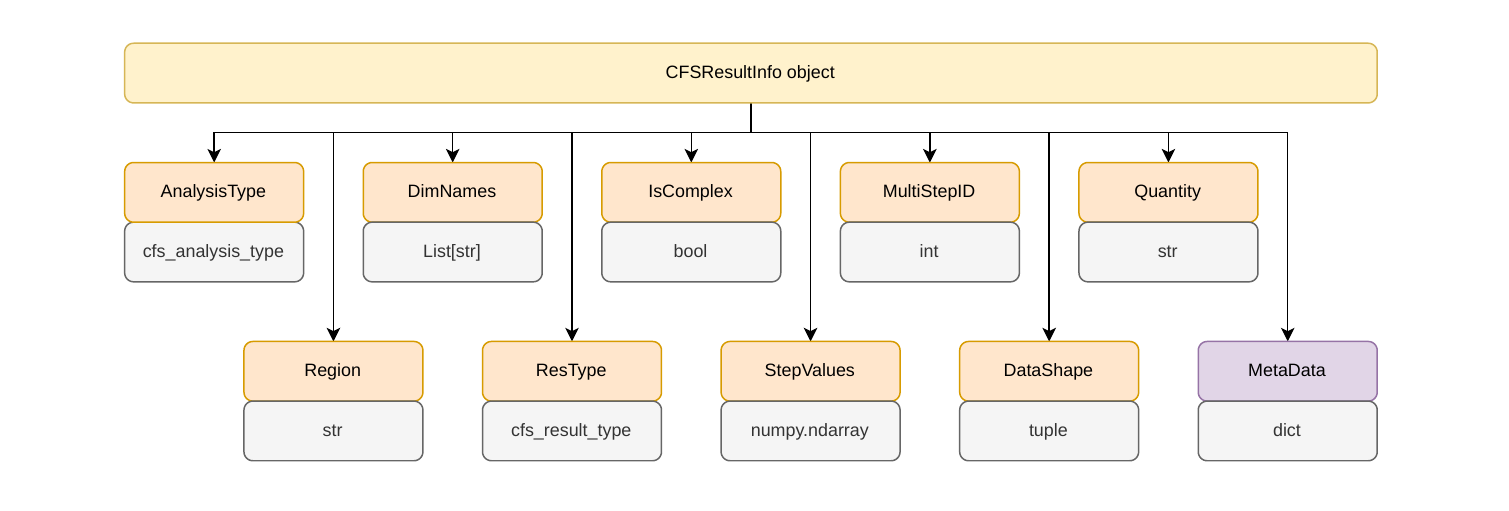}
		\caption{\texttt{CFSResultInfo}}
		\label{fig:data_structures_CFSResultInfo}
	\end{subfigure}
	\caption{Result data structures. Attributes are highlighted in orange, properties in purple, properties with a setter in an orange-to-purple gradient, and respective data types in grey.}
	\label{fig:data_structures_result}
\end{figure}

The \texttt{io} submodule reads and writes openCFS \texttt{.cfs} HDF5 files. It holds their content in the data structures listed in~\cref{tab:data_structures}: the \texttt{CFSMeshData} holds the mesh with its regions, nodes, and connectivity, while \texttt{CFSResultContainer} holds the results of one \texttt{MultiStep} as a list of \texttt{CFSResultArray} items. Each \texttt{CFSResultArray} holds the data and metadata of one quantity on one region. Field data is defined on the nodes or the elements of a region, history data can also be defined on the region itself.
The data structures for meshes and results are illustrated in more detail in~\cref{fig:data_structures_mesh,fig:data_structures_result}, respectively.
Data can be defined on multiple domains, e.g., \texttt{transient}, \texttt{harmonic}, or \texttt{static}, based on the \textbf{AnalysisType} attribute.
The attributes \textbf{DimNames} and \textbf{IsComplex} are set automatically from the shape of the data and the analysis type, unless they are given explicitly. They specify the dimension names of the scalar/vectorial/tensorial data and whether the data is written complex-valued, respectively.
The \textbf{Quantity} attribute can be any string. However, if the data is the input of a subsequent openCFS simulation, the quantity names registered in openCFS must be used, e.g., \texttt{acouRhsLoad} for nodal loads of the acoustic PDE. The registered names are listed in the openCFS XML schema files\footnote{\url{https://gitlab.com/openCFS/cfs/-/tree/master/share/xml/CFS-Simulation/Schemas}}, cf.~\cite{OpenCFSGitLab2026}.
\begin{table}[htbp!]
	\centering
	\caption{Data structures of the \texttt{io} submodule, corresponding to the attributes and properties in~\cref{fig:data_structures_mesh,fig:data_structures_result}.}
	\label{tab:data_structures}
	\small
	\begin{tabularx}{\textwidth}{@{}l>{\raggedright\arraybackslash}p{0.28\textwidth}X@{}}
		\toprule
		Class                       & Description                         & Content                                                                                                                                           \\
		\midrule
		\multicolumn{3}{@{}l}{\textbf{Mesh data structures}}                                                                                                                                                                  \\
		\texttt{CFSMeshData}        & Mesh                                & Node coordinates, element connectivity and types, list of regions, and mesh information                                                           \\
		\texttt{CFSMeshInfo}        & Mesh information                    & Dimension, numbers of nodes and elements per dimension and per element type, quadratic present or not                                             \\
		\texttt{CFSRegData}         & Mesh region                         & Name, node and element IDs, dimension, mesh group or not                                                                                          \\
		\midrule
		\multicolumn{3}{@{}l}{\textbf{Result data structures}}                                                                                                                                                                \\
		\texttt{CFSResultContainer} & Results of one MultiStep            & List of \texttt{CFSResultArray} objects with common analysis type, MultiStep ID, and step values                                                  \\
		\texttt{CFSResultArray}     & Result and metadata                 & Subclass of \texttt{numpy.ndarray}, holding the results of one quantity on one region with metadata.                                              \\
		\texttt{CFSResultInfo}      & Metadata of \texttt{CFSResultArray} & Including the data shape, quantity, region, result type (nodes, elements, region), step values, analysis type, dimension names, and complex flag. \\
		\bottomrule
	\end{tabularx}
\end{table}
\begin{figure}[htb!]
	\centering
	\includegraphics[width=\textwidth]{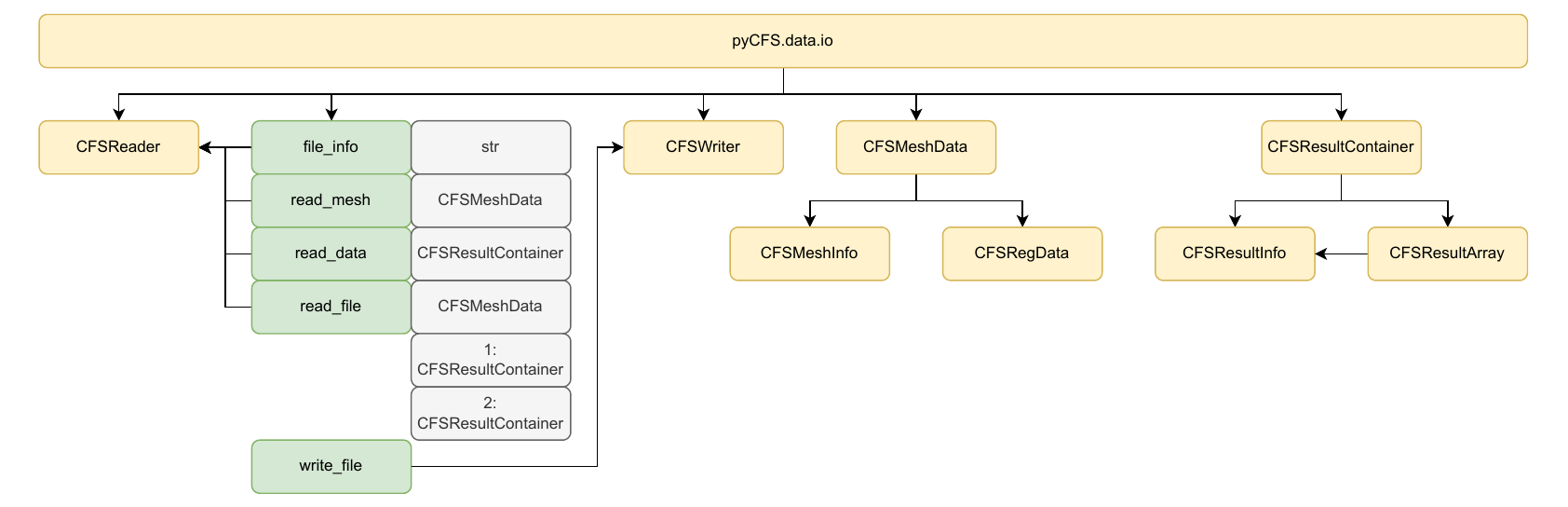}
	\caption{Organization of classes for data structures.}
	\label{fig:data_structures_overview}
\end{figure}

\FloatBarrier
\subsubsection{Reading and writing}

The simplest way of reading and writing files is via the top level utility functions \texttt{read\_file}, \texttt{read\_mesh}, \texttt{read\_data}, \texttt{write\_file}.
A simple example for reading a file and writing the content to a new file is:
\begin{lstlisting}[language=Python]
from pyCFS.data import io

mesh, results = io.read_file('file.cfs')
io.write_file('file_out.cfs', mesh, results)
\end{lstlisting}
Here, \texttt{results} holds a \texttt{CFSResultContainer} per MultiStep. The functions build on the reader and writer classes.

The \texttt{CFSReader} and \texttt{CFSWriter}
\begin{lstlisting}[language=Python]
from pyCFS.data.io import CFSReader, CFSWriter

with CFSReader('file_in.cfs', multistep_id=1) as f:
    mesh = f.MeshData
    results = f.MultiStepData
with CFSWriter('file_out.cfs') as f:
    f.create_file(mesh=mesh, result=results)
\end{lstlisting}
read the file \texttt{file\_in.cfs} into the data objects \texttt{mesh} and \texttt{results}, containing the description of the computational grid and all result data of the MultiStep with the ID given by \texttt{multistep\_id}, respectively, and write this information to a newly created file \texttt{file\_out.cfs}.
By default, the writer compresses data with GZIP level~1 and the HDF5 shuffle filter, optionally with multithreading.

\subsubsection{Partial and chunked access}
For data sets larger than the main memory, \texttt{CFSReader.get\_data\_steps} enables reading selected steps with specified \texttt{step\_numbers}, \texttt{get\_data\_shape} returns data shapes without reading the data, and \texttt{CFSWriter.perform\_write\_multistep} allows writing a chunk of steps into a prepared MultiStep via \texttt{step\_offset}.

\subsubsection{Data at points}
\texttt{CFSReader.get\_data\_at\_points} extracts the results at a set of coordinates, e.g., at microphone or sensor positions. Each coordinate is snapped to the nearest node, or element centroid, of the regions on which the quantity is defined. The data is not interpolated, and a warning is given if the distance exceeds the tolerance \texttt{eps}. Each data point is returned on a point region of its own, e.g., \texttt{Node\_for\_Coord\_0\_on\_fluid}. The corresponding point mesh can be returned by setting \texttt{return\_mesh\_data}.

\subsection{Data processing operators (\texttt{operators})}
\label{sec:operators}
The \texttt{operators} submodule is a library of modules to perform various operations on objects in pyCFS-data data structures, cf.~\cref{sec:io}.
\Cref{tab:operators} lists the main functions with the corresponding pipeline operators in five clusters, which are outlined in the sections below: \texttt{interpolators} in~\cref{sec:interpolation}, \texttt{derivatives} in~\cref{sec:derivatives}, \texttt{timeproc} in~\cref{sec:timeproc}, \texttt{transformation} in~\cref{sec:transformation}, and further operators in~\cref{sec:further_operators}.
\begin{table}[htbp!]
	\centering
	\caption{Main functions of the \texttt{operators} submodule and the corresponding filters of the pipeline.}
	\label{tab:operators}
	\scriptsize
	\begin{tabularx}{\textwidth}{@{}>{\raggedright\arraybackslash}p{3.9cm}>{\raggedright\arraybackslash}X>{\raggedright\arraybackslash}p{3.8cm}@{}}
		\toprule
		Function                                                       & Description                                                                                          & Pipeline filter                           \\
		\midrule
		\multicolumn{3}{@{}l}{\textit{Interpolation} -- \texttt{interpolators}   }                                                                                                                                        \\
		\texttt{interpolate\_node\_to\_cell}                           & mean of the nodal field over each element                                                            & \texttt{Node2Cell}                        \\
		\texttt{interpolate\_cell\_to\_node}                           & volume-weighted mean or least squares fit of element values at the nodes                             & \texttt{Cell2Node}                        \\
		\texttt{interpolate\_nearest\_neighbor}                        & inverse distance weighting of the nearest source points                                              & \texttt{Interpolation\-Nearest\-Neighbor} \\
		\texttt{interpolate\_conservative\_\allowbreak cell\_centroid} & element values as nodal loads, keeping integral and first moment                                     & \texttt{Interpolation\-Cell\-Centroid}    \\
		\texttt{interpolate\_projection\_based}                        & projection onto a source surface, linear FE basis functions                                          & --                                        \\
		\texttt{interpolate\_rbf}                                      & global radial basis function interpolation                                                           & --                                        \\
		\texttt{interpolate\_rbf\_local}                               & local radial basis function interpolation on neighboring points                                      & --                                        \\
		\midrule
		\multicolumn{3}{@{}l}{\textit{Spatial derivatives} -- \texttt{derivatives} }                                                                                                                                      \\
		\texttt{gradient}                                              & gradient or Jacobian, finite element or least squares                                                & \texttt{Gradient}                         \\
		\texttt{divergence}                                            & divergence, finite element or least squares                                                          & \texttt{Divergence}                       \\
		\texttt{curl}                                                  & curl, finite element or least squares                                                                & \texttt{Curl}                             \\
		\texttt{convective\_derivative}                                & $(\bm{u} \cdot \nabla)\varphi$, also on rotating regions                                             & \texttt{Convective\-Derivative}           \\
		\texttt{substantial\_derivative}                               & $\partial \varphi / \partial t + (\bm{u} \cdot \nabla)\varphi$                                       & \texttt{Substantial\-Derivative}          \\
		\texttt{frames} (\texttt{RigidRotation} class)                 & rotating reference frame of a region, taken into account by the derivatives and time averages        & \texttt{ReferenceFrame}                   \\
		\midrule
		\multicolumn{3}{@{}l}{\textit{Time processing} -- \texttt{timeproc}}                                                                                                                                              \\
		\texttt{time\_derivative}                                      & noise-robust five-point differentiator                                                               & \texttt{TimeDerivative}                   \\
		\texttt{time\_average}                                         & mean over all steps or a moving window                                                               & \texttt{TimeAverage}                      \\
		\texttt{time\_rms}                                             & root mean square, by default of the fluctuations                                                     & \texttt{TimeRMS}                          \\
		--                                                             & shift of the step values                                                                             & \texttt{ShiftTime}                        \\
		\midrule
		\multicolumn{3}{@{}l}{\textit{Transformation} -- \texttt{transformation}}                                                                                                                                         \\
		\texttt{transform\_mesh}                                       & rigid body transformation of a mesh and its vector and tensor data                                   & --                                        \\
		\texttt{fit\_mesh}                                             & rigid body fit of regions to a target mesh                                                           & --                                        \\
		\texttt{extrude\_mesh\_region}                                 & 3D or 2D mesh by sweeping a 2D or 1D region along a vector                                           & --                                        \\
		\texttt{revolve\_mesh\_region}                                 & 3D or 2D mesh by revolving a 2D or 1D region about an axis                                           & --                                        \\
		\midrule
		\multicolumn{3}{@{}l}{\textit{Further operators}}                                                                                                                                                                 \\
		\texttt{modal\_analysis}                                       & MAC, MSF, and MCF of mode shapes                                                                     & --                                        \\
		\texttt{field\_fft}                                            & FFT of field data along an axis                                                                      & --                                        \\
		\texttt{dynamic\_mode\_\allowbreak decomposition}              & DMD of field data (PyDMD)                                                                            & --                                        \\
		\texttt{sngr}                                                  & synthetic flow-acoustic sources (SNGR)                                                               & --                                        \\
		\texttt{evaluate\_expression}                                  & math expressions of fields, coordinates, and time. Following the syntax of openCFS math expressions. & \texttt{MathExpression}                   \\
		\bottomrule
	\end{tabularx}
\end{table}

\subsubsection{Interpolation (\texttt{interpolators})}
\label{sec:interpolation}
The \texttt{interpolators} module contains methods for interpolation operations.
The functions and the corresponding pipeline filters are listed in~\cref{tab:operators}.

\begin{figure}[ht!]
	\centering
	\includegraphics{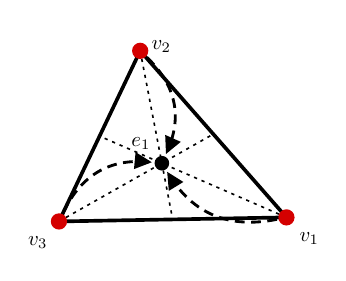} 
	\caption{Node to cell interpolator.}
	\label{fig:interpolators_n2c}
\end{figure}

\paragraph{Node-to-cell interpolation}
The node-to-cell interpolation operator takes nodal values and assigns the mean of the interpolated field over the element to the cell
\begin{equation}
	e_{\square} = \frac{1}{V_\square} \int_{\Omega_\square} \sum_{i=1}^n N_i v_i \,\mathrm{d}V = \sum_{i=1}^n \bigg( \frac{1}{V_\square} \int_{\Omega_\square} N_i \,\mathrm{d}V \bigg) v_i \, .
\end{equation}
Thereby, $e_\square$ is the data assigned to the element, $n$ is the number of nodes of the element, $v_i$ are the nodal values, $N_i$ the FE shape functions, and $V_\square$ the volume of the element $\Omega_\square$. The weights are computed by Gauss quadrature of the shape functions over the element. For linear simplices, i.e., LINE2, TRIA3, and TET4 elements, they equal the arithmetic mean of the nodal values, $1/n$. They differ for pyramids, distorted elements, and quadratic elements, whose corner weights are negative.
The sketch in~\cref{fig:interpolators_n2c} illustrates the Node2Cell operation on one linear triangular element.

\begin{figure}[ht!]
	\centering
	\begin{subfigure}[b]{0.45\textwidth}
		\centering
		\includegraphics{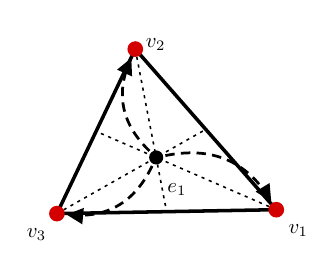} 
		\caption{Element view: each element value $e_i$ contributes to the nodes of the element.}
	\end{subfigure}\hfill
	\begin{subfigure}[b]{0.45\textwidth}
		\centering
		\includegraphics{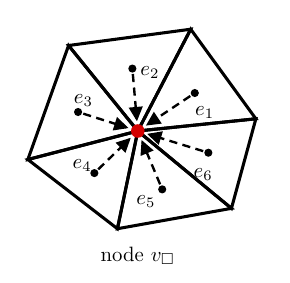} 
		\caption{Node view: the values $e_i$ at the volume centroids $\bm{c}_i$ of the elements around the node are fitted and evaluated at the node.}
	\end{subfigure}
	\caption{Cell to node interpolator.}
	\label{fig:interpolators_c2n}
\end{figure}

\paragraph{Cell-to-node interpolation}
The cell-to-node interpolation operator takes element values, located at the element volume centroids $\bm{c}_i$, and evaluates them at the nodes. With \texttt{degree=0}, the node value is the volume-weighted mean of the adjacent elements of the highest dimension,
\begin{equation}
	v_{\square} = \frac{\sum_{i=1}^n V_i e_i}{\sum_{i=1}^n V_i} \, .
\end{equation}
Thereby, $v_{\square}$ is the data located at the node, $n$ the number of adjacent elements to the node, $e_i$ the data of cell $i$, and $V_i$ its volume.
Note: a mean is exact for constant fields only, and a boundary node has elements on one side only. With the default \texttt{degree=1}, or \texttt{degree=2}, a polynomial $q$ of this degree is therefore fitted by weighted least squares to the values at the $k = 2 n_\phi$ nearest element centroids around the node $\bm{x}_\square$, for $n_\phi$ non-constant monomials of $q$, plus all centroids at the same distance:
\begin{equation}
	\min_q \sum_{i=1}^{k} \frac{\big(q(\bm{c}_i) - e_i\big)^2}{|\bm{c}_i - \bm{x}_\square|^2} \, , \qquad v_\square = q(\bm{x}_\square) \, .
\end{equation}
This is exact for linear fields, or quadratic fields at \texttt{degree=2}, and also at boundary nodes. In a single layer of elements, whose centroids lie in a surface, the fit cannot vary across the layer, which gives a warning. The elements of a stencil must be connected to the node, and singular stencils fall back to a lower-degree polynomial.
The sketches in~\cref{fig:interpolators_c2n} illustrate the element view and the node view of the interpolation.

\paragraph{Nearest neighbor interpolation}

The implemented nearest neighbor interpolation operator makes use of multiple neighbors with inverse distance weighting, i.e., Shepard's method~\cite{shepardTwodimensionalInterpolationFunction1968} with the weights of its modification by Franke and Nielson~\cite{frankeSmoothInterpolationLarge1980}.
Based on the defined number of neighbors $n$ from the source mesh, the nearest neighbors of each target point are searched, and their distances $r_i$ to the considered point are computed.
Based on these distances the weights $w_i$ are computed as
\begin{equation}
	w_i = \left( \frac{R_{\mathrm{ max}}-r_i}{R_{\mathrm{ max}} r_i} \right)^p \, ,
\end{equation}
with $R_\mathrm{ max} = 1.01 r_{\mathrm{ max}}$ as being 1.01 times the maximal distance $r_{\mathrm{ max}}$ of the neighbors, and $p$ as the interpolation exponent.
Finally, each value $v_i$ of the neighbors that is taken into account is weighted to compute the new value $v_{\mathrm n}$ of the considered target point,
\begin{equation}
	v_{\mathrm n} = \sum_{i=1}^{n} \frac{w_i v_i }{\sum_{j=1}^{n}w_{j}} \, .
\end{equation}

\begin{figure}[ht!]
	\centering
	\includegraphics{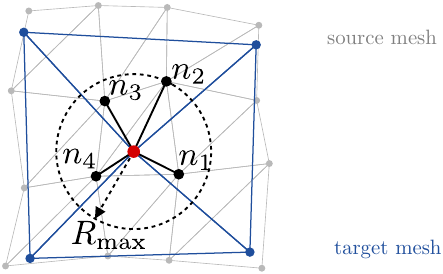} 
	\caption{Nearest neighbor interpolator, with source and target mesh: the values at a target node (red) are interpolated from $n=4$ nearest source nodes, with $R_\mathrm{max}$ derived from the farthest of them.}
	\label{fig:interpolators_nn}
\end{figure}

We recommend setting $1 \leq p \leq 3$, where increasing $p$ shifts weight to the nearer neighbors: $p \to 0$ gives the arithmetic mean of the neighbors, $p \to \infty$ the value of the nearest one. The default is $p=2$, $n=20$.
The weights of each target point sum up to one, so constant fields are reproduced. A target point that coincides with a source point takes its value, and source points with identical coordinates, e.g., a node shared by two source regions, are merged into one point with their mean value.
With \texttt{max\_distance}, neighbors farther away than the specified distance are dropped. The weights are computed from the remaining ones, while target points without remaining neighbors are set to zero.
\Cref{fig:interpolators_nn} illustrates the nearest neighbor interpolation.

\begin{figure}[ht!]
	\centering
	\includegraphics[width=0.45\textwidth]{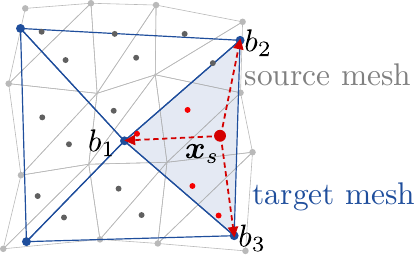}
	\caption{Conservative cell-centroid interpolator, with source and target mesh: the centroids $\bm{x}_s$ (red) of the source cells lie in a target element (blue), whose nodes receive the volume-weighted loads of the source cells, interpolated to the nodes of the target element by the FE shape functions. Each source-cell centroid only contributes to the element containing it.}
	\label{fig:interpolators_conservative}
\end{figure}

\paragraph{Conservative cell centroid interpolation}
\label{sec:conservative}
Source terms of flow simulations are often stored on cell values, whereas an FE simulation needs nodal loads, usually on a different grid.
In the conservative cell-centroid interpolation~\cite{schoderApplicationLimitsConservative2021} \texttt{interpolate\_conservative\_cell\_centroid}, each source cell $s$ with volume-density value $f_s$, volume $V_s$, and volume centroid $\bm{x}_s$ contributes to the nodes $a$ of the target element that contains $\bm{x}_s$,
\begin{equation}
	b_a = \sum_{s} f_s V_s N_a\big(\bm{\xi}(\bm{x}_s)\big) \, ,
\end{equation}
with the FE shape functions $N_a$ of the target element at the local coordinates $\bm{\xi}(\bm{x}_s)$ corresponding to the source-cell centroid. Since the shape functions sum up to $1$ and interpolate within the element via $\sum_a N_a \bm{x}_a = \bm{x}$, the loads keep the approximated integral $\sum_a b_a = \sum_s f_s V_s$ and the first moment $\sum_a b_a \bm{x}_a = \sum_s f_s V_s \bm{x}_s$ of the source field, summed over all source cells whose centroids lie in the target mesh.
All source-cell centroids outside the target mesh are dropped, with a warning.
Local coordinates slightly outside an element are projected onto it.
The interpolation is supported on linear and quadratic elements, including surface regions in 3D.
\Cref{fig:interpolators_conservative} illustrates the conservative cell-centroid interpolation.

\paragraph{Projection-based interpolation}

\begin{figure}[ht!]
	\centering
	\includegraphics[width=0.8\textwidth]{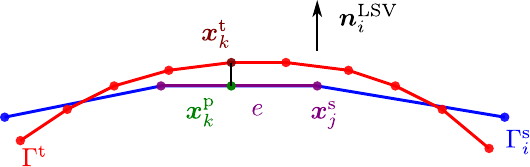}
	\caption{Projection-based interpolation. Sketch of an arbitrary source (blue) and target mesh (red) with indicated point projection from the target to the source mesh, from\cite{wurzingerExperimentalPredictionMethod2024}.}
	\label{fig:interpolators_projection_interpolation}
\end{figure}

This interpolation operator projects points of the target mesh onto the source mesh and evaluates the result based on linear FE basis functions. A comprehensive description of the algorithm can be found in~\cite{wurzingerExperimentalPredictionMethod2024}.
\Cref{fig:interpolators_projection_interpolation} illustrates the projection-based interpolation.
Available parameters of the method include
\begin{itemize}
	\item \texttt{projection\_direction}: Direction vector used for projection. It can be specified as constant or individually for each node. By default, the node normal vector (based on averaged neighboring element normal vectors) is used.
	\item \texttt{max\_projection\_distance}: Maximum projection distance. Lower values speed up interpolation matrix build and prevent projecting onto far surfaces.
	\item \texttt{search\_radius}: Source elements farther than this radius from the target point are not considered as projection targets. It should be chosen at least as large as the maximum element size of the source grid (default: \texttt{max\_projection\_distance}).
\end{itemize}

\paragraph{Radial basis function interpolation}
\label{sec:rbf}
This module provides functions to interpolate data using radial basis functions (RBF). Implemented kernel functions are: Gaussian RBF kernel, multiquadric RBF kernel, and Wendland C2 RBF kernel.

The implementation features the interpolation of a field quantity using global or local RBF approximation. Parameters include
\begin{itemize}
	\item \texttt{kernel}: Type of RBF kernel to use. Available kernels include \texttt{gaussian}, \texttt{multiquadratic}, and \texttt{wendland\_c2}.
	\item \texttt{epsilon}: Shape parameter for the used RBF kernel function (the support radius for \texttt{wendland\_c2})
	\item \texttt{smoothing}: Smoothing parameter for the kernel matrix
\end{itemize}
for the global interpolation and additionally for the local interpolation
\begin{itemize}
	\item \texttt{neighbors}: Number of nearest neighbors to use for local interpolation
	\item \texttt{min\_neighbors}: Minimum number $k_\mathrm{min}$ of neighbors to use based on a local neighborhood radius \mbox{$r = f_r \frac{1}{k_\mathrm{min}} \sum_{i=1}^{k_\mathrm{min}} d_i$} with factor $f_r$, the distances $d_i$ of the target point to its $k_\mathrm{min}$ nearest source points.
	      Of the \texttt{neighbors} nearest points, those within $r$, but at least $k_\mathrm{min}$, are used. If \texttt{min\_neighbors} is not set (default), all \texttt{neighbors} nearest points are used.
	\item \texttt{radius\_factor}: Factor $f_r$ to determine the local neighborhood radius.
\end{itemize}

\subsubsection{Spatial derivatives (\texttt{derivatives})}
\label{sec:derivatives}

The functions \texttt{gradient}, \texttt{divergence}, and \texttt{curl} differentiate node or element data of a region with a sparse matrix $\mathbf{G}_d$ per coordinate direction $d$. The matrix is built once per region, so stencils do not cross region boundaries and discontinuities at material interfaces stay sharp.
The functions and the corresponding pipeline filters are listed in~\cref{tab:operators}.
Two gradient-computation methods are currently implemented. Meshes with 3D coordinates in a plane $z = \mathrm{const}$ give two-dimensional derivatives.

\paragraph{Finite element gradient}
With \texttt{method="fe"}, for node data, the gradient of each element is the volume mean of the gradient of its Lagrange interpolation,
\begin{equation}
	\overline{\nabla u}_e = \frac{1}{V_e} \int_{\Omega_e} \nabla \sum_{a} N_a u_a \,\mathrm{d}V \, ,
\end{equation}
for all linear and quadratic Lagrange elements.
The gradient of the interpolation is constant on linear simplices, but varies within the element for other elements.
For quadratic fields on straight-sided quadratic elements, the volume mean equals the gradient at the volume centroid of the element.
The gradient at the nodes is derived as the volume-weighted mean of the adjacent elements for linear elements, or, at nodes whose adjacent elements are all quadratic volume elements, the linear fit of the element gradients at their volume centroids.
The method is exact for linear fields and for quadratic fields on straight-sided quadratic elements.

\paragraph{Least squares gradient}
With \texttt{method="lsq"}, for node or element data, a polynomial of degree 1 or 2 is fitted to the values $u_j$ at the $k$ nearest data points $\bm{x}_j$ around each evaluation point $\bm{x}_i$.
The fitted polynomial is a local Taylor expansion around $\bm{x}_i$. If $\bm{x}_i$ is itself a data point, $c_0$ is fixed to $u_i$, so that the fit passes through its value:
\begin{equation}
	\min_{\bm{c}} \sum_{j=1}^{k} w_j \Big( c_0 + \sum_{m} c_m \phi_m(\bm{s}_j) - u_j \Big)^2 , \qquad \bm{s}_j = \frac{\bm{x}_j - \bm{x}_i}{h} , \qquad w_j = \frac{h^2}{|\bm{x}_j - \bm{x}_i|^2} \, ,
\end{equation}
with the non-constant monomials $\phi_m$ of the fitted polynomial, and the stencil radius $h = \max_j |\bm{x}_j - \bm{x}_i|$.
The gradient is given by the linear coefficients divided by $h$.
By default, a stencil holds $k = 2 n_\phi + 1$ points for $n_\phi$ non-constant monomials. For example, $k = 7$ for degree 1 in 3D, plus all further points at the same distance as the $k$-th one, so that stencils on structured grids are symmetric.
Singular stencils, where points don't provide sufficient information to determine all monomials, and degenerate stencils, where a direction is only resolved by a small spread of points, are treated specifically.
They are extended by doubling the number of neighbors, up to a maximum number. Degree 2 stencils that remain singular, e.g., in layers of less than three points, are fitted with degree 1. Degenerate stencils that still resolve too few directions get only the derivatives along the resolved ones, with a warning.
The fit is exact for polynomials of its degree and works on any point cloud, e.g., the cells of polyhedral finite volume meshes.

\paragraph{Surface gradients}
On surface regions in 3D and line regions in a 2D setup, both methods return the tangential gradient, with
\begin{equation}
	\nabla_\Gamma N_a = \mathbf{J}(\mathbf{J}^\mathsf{T}\mathbf{J})^{-1}\nabla_{\bm{\xi}} N_a.
\end{equation}
The Jacobian $\mathbf{J} = \partial \bm{x} / \partial \bm{\xi}$ of the element is taken from the FE shape functions for \texttt{method="fe"}, or a height function fit to the normal offsets of the points of the surface, for \texttt{method="lsq"}.

\paragraph{Divergence and curl}
The \texttt{divergence} $\nabla \cdot \bm{u}$ and the \texttt{curl} $\nabla \times \bm{u}$ functions use the matrices of the chosen gradient-computation method. The curl is not available on surface and line regions.

\paragraph{Convective and substantial derivative}
The function \texttt{convective\_derivative} and \texttt{substantial\_derivative} compute
\begin{equation}
	(\bm{u} \cdot \nabla)\varphi \qquad \text{and} \qquad \frac{\mathrm{D}\varphi}{\mathrm{D}t} = \frac{\partial \varphi}{\partial t} + (\bm{u} \cdot \nabla)\varphi,
\end{equation}
respectively. They operate on scalar or vectorial data $\varphi$.
The time derivative for the substantial derivative is computed as described in~\cref{sec:time_derivative}, and requires equidistant steps.
The velocity $\bm{u} = \bm{u}_\mathrm{f} + \bm{f}(\bm{x}, t)$ is either read from the field data, $\bm{u}_\mathrm{f}$, or specified as an analytical function, $\bm{f}(\bm{x}, t)$. If both are given, their superposition is applied.
A moving reference frame can be specified, which sets $\bm{f}(\bm{x}, t) = -\bm{u}_{\mathrm{grid}}$, i.e., applies an Arbitrary Lagrangian-Eulerian (ALE) reference frame~\cite{doneaChapter14Arbitrary2004}.
The velocity field may also be time-constant, e.g., a mean velocity, cf.~\cref{sec:example}.

\paragraph{Rotating reference frames}
The flow in a rotating part, e.g., an impeller, is often exported on a static mesh of the part at a reference time $t_\mathrm{ref}$, with vectors in global orientation.
In this case, a mesh point $\bm{X}$ is at $\bm{x}_0 + \mathbf{R}(t)(\bm{X} - \bm{x}_0)$ at the time $t$, with the rotation $\mathbf{R}(t)$ by the angle $\omega (t - t_\mathrm{ref})$ about the axis $\hat{\bm{a}}$ through $\bm{x}_0$, specified by a \texttt{RigidRotation} frame with the rotational speed in rpm, $\omega = 2 \pi\, \mathrm{rpm} / 60$. For such a moving reference frame, the \texttt{frames} argument of the operators rotates vectors into the frame, $\bm{v}' = \mathbf{R}(t)^\mathsf{T} \bm{v}$, before they are averaged in time or spatially differentiated. This way, the mean circumferential velocity does not cancel. The convective and substantial derivatives use the velocity relative to the moving mesh, $\mathbf{R}(t)^\mathsf{T} \bm{u} - \omega \hat{\bm{a}} \times (\bm{X} - \bm{x}_0)$.
In a pipeline, the \texttt{ReferenceFrame} filter declares the frames once for all downstream filters.

\subsubsection{Time signal processing (\texttt{timeproc})}
\label{sec:timeproc}

The \texttt{timeproc} module contains methods for time domain data processing operations.
The functions and the corresponding pipeline filters are listed in~\cref{tab:operators}.

\paragraph{Time derivative}
\label{sec:time_derivative}

The time derivative of the desired quantity $\partial q / \partial t$ can be computed using \texttt{timeproc.time\_derivative}, which applies a noise-robust second-order differentiation scheme with an $N=5$ point stencil. The scheme suppresses  high frequencies and is precise on low frequencies, according to~\cite{SmoothNoiserobustDifferentiators}.
The time derivative is calculated by
\begin{equation}
	\frac{\partial q}{\partial t} \approx \frac{2(q_1-q_{-1})+q_2-q_{-2}}{8 \Delta t} \,,
\end{equation}
where the index of $q$ defines the time step relative to the time step of which the derivative is calculated and $\Delta t$ is the time step size, which must be equidistant and is checked with a relative tolerance of $10^{-6}$.

Parameters include
\begin{itemize}
	\item \texttt{boundary\_treatment}: The start and end of the time series can either be removed (\texttt{remove}), left untreated (\texttt{None}), where the two steps at each end are not valid derivatives and a warning is issued, or treated by one-sided estimators according to \cite{SmoothNoiserobustDifferentiators} (\texttt{one-sided}), which are first-order accurate at these steps. At least 5, 2, or 7 steps are required, respectively. Harmonic and eigenvalue analyses, whose steps are not times, are rejected.
\end{itemize}

\paragraph{Time average and root mean square (RMS)}
\label{sec:time_average}
The time average is computed by \texttt{timeproc.time\_average} as the arithmetic mean of the steps $q_m$, either over all steps or over a moving window of $w$ steps, which can be centered, for odd $w$, or causal,
\begin{equation}
	\bar{q}_n = \frac{1}{w} \sum_{m = n - (w-1)/2}^{n + (w-1)/2} q_m \qquad \text{or} \qquad \bar{q}_n = \frac{1}{w} \sum_{m = n-w+1}^{n} q_m \, .
\end{equation}
Steps whose window extends beyond the start or end of the time series are either removed or averaged over the truncated window.
For non-equidistant steps, the arithmetic mean differs from the time-weighted mean, in which case a warning is printed.
The root mean square (RMS) is similarly computed by \texttt{timeproc.time\_rms}.
The RMS is by default computed as standard deviation around the mean of the window, and is optionally computed as absolute RMS, i.e.,
\begin{equation}
	\sqrt{\overline{|q - \bar{q}|^2}}, \qquad \text{and} \qquad \sqrt{\overline{|q|^2}},
\end{equation}
respectively.
For vectorial data, time average and RMS are computed per component. The pipeline filter \texttt{ShiftTime} shifts the step values by a constant offset.
Vectors of rotating regions can be averaged in frame orientation to avoid cancellation of the circumferential components, cf.~\cref{sec:derivatives}.

\subsubsection{Mesh transformation (\texttt{transformation})}
\label{sec:transformation}
The \texttt{transformation} module contains methods for mesh/data transformation operations.
The functions are listed in~\cref{tab:operators}.

\paragraph{Mesh extrusion and revolution}

The functions \texttt{extrude\_mesh\_region} and \texttt{revolve\_mesh\_region} create a 3D/2D mesh by extruding/revolving a 2D/1D base mesh along a vector or about an axis through the origin, respectively, in a given number of layers. Line, triangle, and quadrilateral elements (LINE2, TRIA3, QUAD4) become QUAD4, WEDGE6, and HEXA8 elements. A full revolution closes the mesh.

\paragraph{Rigid body transformation}

The function \texttt{transform\_mesh\_data} performs a rigid body transformation of a mesh. Vector and tensor data are rotated with the mesh by \texttt{transform\_result\_data}, as $\mathbf{R}\bm{v}$ and $\mathbf{R}\mathbf{T}\mathbf{R}^\mathsf{T}$, respectively. The mesh and its data together are transformed by \texttt{transform\_mesh}.
Symmetric tensors are stored in the Voigt order of openCFS (xx, yy, zz, yz, xz, xy). 
Note that strains hold the engineering shear strains and stresses the tensor components. 
Hence, the strain quantities must be named by \texttt{strain\_quantities}.
Two-dimensional data is rotated in the $xy$ plane, and a rotation out of this plane is rejected before the mesh or the results are changed. Other non-scalar data is left unchanged with a warning.
The transformation is given by a translation $\bm{t}$ and three angles $(\alpha, \beta, \gamma)$ in rad of rotations about the fixed global $x$, $y$, and $z$ axes, through an origin $\bm{x}_0$ given by \texttt{rotate\_origin}.
The angles are specified as the extrinsic Euler angles, $\mathbf{R} = \mathbf{R}_z(\gamma)\mathbf{R}_y(\beta)\mathbf{R}_x(\alpha)$. The transformation can be restricted to selected regions.

\paragraph{Mesh fit}
The function \texttt{fit\_mesh} fits a region to a target region using rotation and translation transformations. The transformation parameters are x,y,z displacement, and Euler angles.
In contrast to \texttt{transform\_mesh}, the rotation is about the global origin.
The objective, the sum of the squared distances of the transformed source nodes to their nearest target nodes, which are found with a k-d tree, is minimized by the default quasi-Newton method of SciPy (BFGS) from an initial guess. For larger initial misalignments, a global differential evolution optimizer with bounds on the translation and the angles can be chosen.
The function returns the transformed mesh and results and the parameters of each fit. Typical applications are the alignment of measurement grids, e.g., of a laser vibrometer with an FE mesh~\cite{wurzingerDevelopmentValidatedSimulation2025,heideggerModeorientedFiniteelementModel2026}. In addition, \texttt{project\_mesh\_onto\_plane} projects regions onto the best-fit plane of a given set of points.

\subsubsection{Further operators}
\label{sec:further_operators}

\paragraph{Modal analysis}
Common metrics used in modal analysis including the Modal Assurance Criterion (MAC), the Modal Scale Factor (MSF), and the Modal Complexity Factor (MCF), are given in the \texttt{modal\_analysis} module.

\paragraph{Fast Fourier transform (FFT)}

The function \texttt{field\_fft} applies the Fast Fourier Transformation (FFT) to the field data along a specified axis.

\paragraph{Dynamic mode decomposition (DMD)}

The function \texttt{dynamic\_mode\_decomposition} applies the Dynamic Mode Decomposition (DMD) to the field data.
This submodule makes use of \textit{pydmd}, which is not a standard requirement of \textit{pyCFS}. The additional dependencies can be installed via pip using \texttt{pip install pycfs[dmd]}.

\paragraph{Synthetic noise generation and radiation (SNGR)}
The module \texttt{sngr} covers the computation of synthetic flow acoustic sources based on the theory of~\cite{StochasticApproachCompute}, for the Synthetic Noise Generation and Radiation (SNGR) with additional description found in \cite{weitzApproachComputeCavity}. The implementation is described in detail in~\cite{wurzingerDevelopmentValidatedSimulation2025,wurzingerHydrodynamicFlowacousticSimulation2025}.

\paragraph{Math expressions}
The function \texttt{evaluate\_expression} of the \texttt{expression} module and the corresponding pipeline filter \texttt{MathExpression} evaluate math expressions on result fields, e.g., scaling, offsets, products of fields, and functions of the time or frequency and of the coordinates.
The expressions are written as strings in the syntax of the openCFS math expressions.
For example, an expression
\begin{lstlisting}[language=Python]
	f"acouPressure*if(t lt {TAU}, 1 - cos(pi*t/(2*{TAU}))^2, 1)"
	\end{lstlisting}
applies a smooth temporal onset to the quantity \texttt{acouPressure}, at all time steps smaller than the onset length \texttt{TAU}.
Variables are the quantity names of the fields, the coordinates \texttt{x}, \texttt{y}, and \texttt{z} of the nodes or element centroids, the time \texttt{t} of transient data or the frequency \texttt{f} of harmonic data, and the constant \texttt{pi}.
The operations act on each component of vector data separately.
An application is shown in~\cref{sec:example}.

\subsection{Data processing pipeline (\texttt{pipeline})}
\label{sec:pipeline}
The operator functions of~\cref{sec:operators} are designed to hold all steps of a result in memory. In contrast, the \texttt{pipeline} submodule applies a chain of operators to one chunk of steps at a time, facilitating operations on large data sets, such as computing aeroacoustic source terms from computational fluid dynamics (CFD) data.
The C++ tool openCFS-data (cfsDat)~\cite{schoderOpenCFSDataDataPrePostProcessing2023} provides similar operations, configured by XML files.
Since version~0.3.0, the pyCFS-data \texttt{pipeline} offers most of cfsDat's functionality and additional, new features.
Like an XML file of cfsDat, a pipeline describes a complete processing chain from the input files to the output files.
It provides comprehensive pre-checks to identify mismatches in the file metadata or the operator chains before the full chain is run.
For transient data sets exceeding a specified memory limit, a \texttt{pipeline} is processed in chunks of data.
With the pipeline, pyCFS-data provides a more flexible alternative to cfsDat for complete processing chains, and it aims to supersede cfsDat in the future.

\begin{figure}[htb!]
	\centering
	\includegraphics[width=0.95\textwidth]{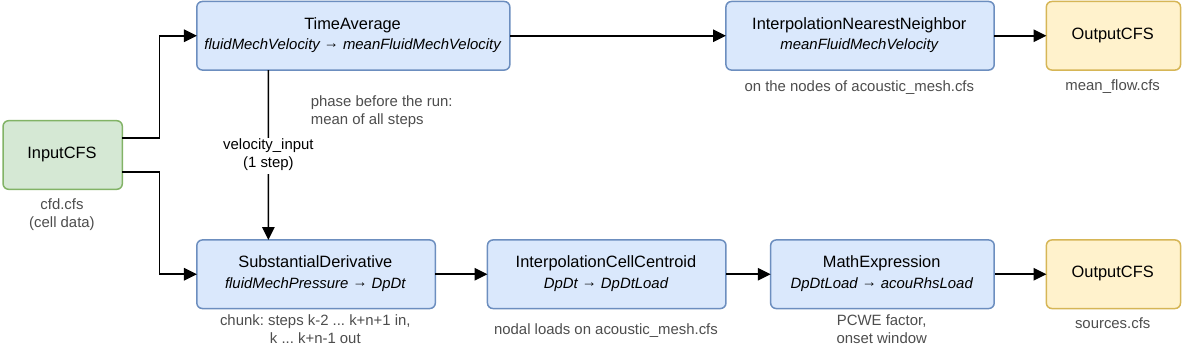}
	\caption{Data processing pipeline corresponding to the example in~\cref{sec:example}: filter graph computing the mean flow and the substantial derivative of CFD data.}
	\label{fig:pipeline}
\end{figure}

\paragraph{Filters and streams}
Operations within the pipeline are referred to as \textit{filters}, including input, operator, and output filters.
Each filter receives its upstream filters as arguments.
The pipeline collects the filter graph from its last filters, as shown in~\cref{fig:pipeline}.
The input filters \texttt{InputCFS} and \texttt{InputEnSight} read openCFS HDF5 files and EnSight Case Gold files, respectively. The output filter \texttt{OutputCFS} writes openCFS HDF5 files. The operator filters are listed in \cref{tab:operators}.
Each filter provides a stream of metadata, i.e., quantities, regions, result types, step values, data shapes, reference frames.
Input and output filters select quantities with \texttt{quantities}, a list of names. Operator filters take \texttt{quantity\_names}, as the operator functions do: a list selects quantities, whose outputs are named by a default suffix, e.g., \texttt{\_mean}, whereas a dict maps the input quantities to the names of the outputs.
During the run, the data of the current chunk are passed as arrays of shape (steps, entities, components).
The graph can branch and merge: a filter can feed several filters, and filters can take further inputs, e.g., the substantial derivative takes the velocity from a \texttt{TimeAverage} filter via \texttt{velocity\_input}, so that the mean velocity is computed within the same pipeline.
In this case, the time averaging is performed in a phase before the main run and used as convective velocity afterward.
Custom filters can be defined, which need to implement \texttt{infer} (metadata) and \texttt{process} (one chunk) of the base class \texttt{Filter}. 
The input arrays of a chunk are read-only, as they are shared between filters and chunks, so a filter returns new arrays.

\paragraph{Dry run}
\texttt{Pipeline.check} propagates the metadata from the file attributes and dataset shapes of the inputs through all filters. Each filter checks its inputs, e.g., element data for the conservative interpolation, and reports all setup errors at once, before any bulk data is read, except for the geometry of EnSight parts.
The check also confirms that no output overwrites an input and estimates the memory and the required chunk size.

\begin{figure}[htb!]
	\centering
	\includegraphics[width=0.95\textwidth]{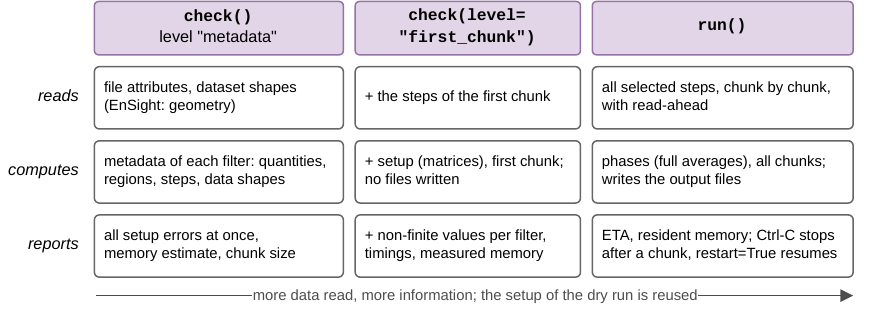}
	\caption{Dry runs and run: what each mode reads, computes, and reports.}
	\label{fig:dry_runs}
\end{figure}

With the argument \texttt{level="first\_chunk"}, the filter matrices are assembled and the first chunk of data is processed, without writing to files.
The \texttt{first\_chunk} dry run reports timings of each filter, the measured memory, and non-finite values where they first appear. The setup is reused by the full run.
\Cref{fig:dry_runs} illustrates the two dry-run modes and the full-run mode.

\paragraph{Chunked execution}
For each chunk of output steps, the pipeline determines backwards from the outputs which steps each filter needs.
For example, the five-point stencil of the time derivative, cf.~\cref{sec:time_derivative}, needs two additional steps on each side.
Steps shared with the previous chunk are kept instead of being read and computed again.
Spatial operators are sparse matrices $\mathbf{A}$, built once in the setup and applied to a chunk by one sparse product.
Temporal operators are sparse matrices along the steps, built for the steps of each chunk.
The average and RMS over all steps are computed in a phase before the main run, which accumulates the deviations from the first step chunk by chunk, $\bar{q} = q_1 + \frac{1}{N} \sum_{m=1}^{N} (q_m - q_1)$, with the memory of a few steps. The filters upstream of such a phase run again in the main run, so the input is read twice. Only the quantities and regions that the filters need are read and computed.

\paragraph{Memory model}
The pipeline estimates the peak memory $P(n)$ of a chunk of $n$ steps from the metadata, by simulating its processing. It considers the data of each filter, stencil halos, temporary arrays, matrices, read-ahead, and writing processes.
The chunk size is the largest $n$ with $s\,P(n) \leq L$ for the memory limit $L$ and the safety factor $s = 1.5$.
After the setup, chunk size is chosen again with the measured size of the matrices.

\paragraph{Concurrency and restart}
Inputs read the next chunk in a separate thread, and \texttt{OutputCFS} writes in a background process with multithreaded compression.
Soft-interrupting is supported similar to openCFS simulations, where the first interrupt via \textit{SIGINT} stops a run after the current chunk with valid output files, or, during a phase, before the output files are created.
Further interrupts are ignored.
A termination via \textit{SIGTERM} or an error stops the run at once, after the writing process has completed the chunk it is writing. The output files then keep the steps of the chunks written before, but are not finalized.
As the last written step is recorded in the files after each chunk, this also holds if the process is killed via \textit{SIGKILL}.
In all these cases, a re-run with \texttt{run(restart=True)} continues after checking the stored steps, the pipeline metadata, the mesh, and a recomputed step.

\subsection{Other data formats (\texttt{extras})}
\label{sec:extras}

The \texttt{extras} module covers libraries of alternative file formats to read from, convert to, and write in.
We recommend EnSight Gold files for CFD data, and \texttt{.cfs} files or, if these cannot be written, Exodus files for finite element meshes.

\subsubsection{EnSight Case Gold}
\label{sec:ensight}
The \texttt{ensight\_io} module contains data processing utilities for reading EnSight Case Gold files, e.g. exported from common CFD software like OpenFOAM, Ansys Fluent, or STAR-CCM+. Since most CFD solvers export this format, and the reader handles polyhedral cells and transient geometry, it is the recommended format for CFD data. The pipeline reads it directly with \texttt{InputEnSight}.

The reader supports conversion of polygons and polyhedra, which openCFS files cannot hold. Cells of a standard shape become standard elements, and the other cells are decomposed around their face and cell centers into triangles, tetrahedra, and pyramids. The decomposition keeps the mesh conforming and the cell volumes exact. Values at the added nodes are interpolated, so that linear fields are reproduced. Element data of a decomposed cell keeps the value of the cell on all its elements.

The \texttt{EnSightCaseReader} reads selected parts, variables, and steps of C binary and ASCII files: static geometry is read once, transient geometry is supported, and the time values are taken from the case file. Parts are merged by their shared nodes, and unreadable or ambiguous parts are reported before any data is read. The data is converted to CFSMeshData and CFSResultContainer objects.
Note that this submodule makes use of \textit{VTK} (versions 9.5 to 9.7 supported), which is not a standard requirement of \textit{pyCFS}. The additional dependencies can be installed via pip using \texttt{pip install pycfs[vtk]}.

\subsubsection{CGNS}
The \texttt{cgns\_io} module reads mesh files in CGNS file format, e.g. exported from Ansys Mechanical.
Note that only HDF5 based CGNS files can be read. Older CGNS versions also utilize the ADF file format, which must be converted to HDF5 beforehand. This can be done e.g. on a Linux system using \texttt{cgnslib} with the command
\texttt{adf2hdf \$CGNS\_FILE}. If \texttt{adf2hdf} is found on the system path, \texttt{read\_mesh} converts ADF files automatically.

\subsubsection{Ansys Mechanical}
The \texttt{ansys\_io} module contains data processing utilities for reading Ansys RST files.
It reads Ansys result files (\texttt{*.rst}) into CFSMeshData and CFSResultContainer objects.
This submodule makes use of the \textit{Ansys Data Processing Framework (DPF)}, which is not a standard requirement of \textit{pyCFS} and requires a licensed installation of the according DPF server. Since DPF is in Beta development stage, compatibility might break rapidly. The additional dependencies can be installed via pip using \texttt{pip install pycfs[ansys]}.

\subsubsection{Exodus}
The \texttt{exodus\_io} module covers reading mesh files in Exodus file format, e.g. generated with Coreform Cubit. Element blocks, node sets, and side sets become regions. Exodus is the recommended format for finite element meshes that cannot be exported as \texttt{.cfs} files directly.

\subsubsection{meshio}
The \texttt{meshio\_io} interface loads meshes via meshio into the pyCFS CFSMeshData format.
This submodule makes use of \textit{meshio}, which is not a standard requirement of \textit{pyCFS}. The additional dependencies can be installed via pip using \texttt{pip install pycfs[meshio]}.

\subsubsection{STL}
The \texttt{stl\_io} module reads mesh data from STL files.

\subsubsection{Polytec PSV}
\label{sec:psv}
The \texttt{psv\_io} module contains data processing utilities for reading universal files (UFF) exported from the PSV Software of Polytec scanning laser Doppler vibrometers.
A typical workflow reads the measurement grid with \texttt{read\_unv\_mesh} and the measured data, e.g., an H1 estimate of the frequency response function, with \texttt{read\_unv\_data\_by\_info}. The function \texttt{check\_unv} lists the data sets in a file. Converting to the \texttt{CFSMeshData} and \texttt{CFSResultArray} objects formats can be done with \texttt{convert\_mesh\_to\_cfs} and \texttt{convert\_data\_to\_cfs}, respectively, and \texttt{convert\_from\_cfs} converts them back.
Since a scanning vibrometer measures the velocity component along its laser beam, \texttt{read\_unv\_mesh} optionally computes the position of the scanning head from measured distances. With it, a single 1D measurement becomes a vector along the beam direction, and \texttt{combine\_3D} combines three measurements from different head positions into 3D vectors. The measurement grid can then be aligned with a finite element mesh by \texttt{fit\_mesh}, cf.~\cref{sec:transformation}, and the data be interpolated onto it, cf.~\cref{sec:interpolation}.
Similar workflows have been successfully applied, e.g., in~\cite{heideggerModeorientedFiniteelementModel2026,wurzingerDevelopmentValidatedSimulation2025,wurzingerExperimentalPredictionMethod2024,HeideggerModelingandSimulation}.

\subsubsection{NiHu}
The \texttt{nihu\_io} module contains data processing utilities for reading NiHu~\cite{fialaNiHuOpenSource2014} Matlab structures.
It converts the NiHu mesh Matlab structure into a CFSMeshData object, and NiHu result data into CFSResultContainer objects.

\section{Illustrative example}
\label{sec:example}
This example computes an aeroacoustic source term of the perturbed convective wave equation (PCWE)~\cite{kaltenbacherComputationalAeroacousticsRotating2017,schoderPerturbedConvectiveWave2026} from transient flow data and maps it onto an acoustic mesh, as illustrated in~\cref{fig:pipeline}.
The right-hand side of the PCWE is proportional to the substantial derivative of the flow pressure $p$ with the mean flow velocity $\bar{\bm{u}}$, $\partial p/\partial t + (\bar{\bm{u}} \cdot \nabla) p$. The source file \texttt{cfd.cfs} holds the pressure and the velocity of $n_t$ time steps as cell data of a finite volume simulation stored on a fine hexahedral grid of $41^3$ nodes. The file \texttt{acoustic\_mesh.cfs} holds a coarser hexahedral grid of $21^3$ nodes. The pipeline averages the velocity over all steps and uses the mean as the convective velocity of the substantial derivative, with least squares gradients of the cell data, cf.~\cref{sec:derivatives}. The conservative interpolation maps the result as nodal loads onto the acoustic mesh, and the mean velocity is interpolated onto its nodes. A \texttt{MathExpression} filter then scales the nodal loads by the PCWE factor $1/(\rho_0 c_0^2)$, with the density $\rho_0$ and the speed of sound $c_0$, and applies a smooth onset window, $1 - \cos^2(\pi t / (2 \tau))$ for $t < \tau$. As both operations are linear, the filter is applied after the interpolation, where it acts on the coarser acoustic mesh.
\begin{lstlisting}[language=Python, basicstyle=\ttfamily\footnotesize, columns=fixed, basewidth=0.525em, keepspaces=true]
from pyCFS.data import pipeline as pl

src = pl.InputCFS("cfd.cfs")
mean = pl.TimeAverage(src, quantity_names={"fluidMechVelocity": "meanFluidMechVelocity"})
dpdt = pl.SubstantialDerivative(src, quantity_names={"fluidMechPressure": "DpDt"},
                                method="lsq", velocity="meanFluidMechVelocity",
                                velocity_input=mean)
regions = {"source": ["fluid"], "target": ["fluid"]}
loads = pl.InterpolationCellCentroid(dpdt, "acoustic_mesh.cfs", regions,
                                     quantity_names={"DpDt": "DpDtLoad"})
rho_0, c_0, tau = 1.204, 343.0, 5e-3  # kg/m^3, m/s, s
window = f"if(t lt {tau}, 1 - cos(pi*t/(2*{tau}))^2, 1)"
source = pl.MathExpression(loads, {"acouRhsLoad": f"DpDtLoad/({rho_0}*{c_0}^2)*{window}"})
mean_acou = pl.InterpolationNearestNeighbor(mean, "acoustic_mesh.cfs", regions)
pipe = pl.Pipeline(pl.OutputCFS(source, "sources.cfs"),
                   pl.OutputCFS(mean_acou, "mean_flow.cfs"),
                   memory_limit=0.2)  # GiB
pipe.check().raise_if_failed()  # dry run: metadata, memory estimate, chunk size
pipe.run()
\end{lstlisting}
The dry run lists the metadata of each filter and the memory estimate. The mean velocity is computed in a phase before the main run, which reads all $n_t=200$ steps in $6$ chunks, and the main run fits chunks of $20$ steps within the limit of \unit[0.2]{GiB}, with an estimated peak of \unit[133]{MiB}. After the setup, the chunk size is chosen again with the measured size of the matrices, so that \texttt{run()} processes the 196 steps of the substantial derivative, without two steps at each end required by the stencil, in 9 chunks of up to 22 steps. The output files hold the quantities \texttt{acouRhsLoad} and \texttt{meanFluidMechVelocity}, which are names registered in openCFS, cf.~\cref{sec:io}.

\section{Impact}
\label{sec:impact}
The pyCFS-data package was used in hybrid experimental-numerical data processing routines to estimate sound power radiation of shell and plate structures including a Gaussian Process Regression~\cite{wurzingerUncertaintyPropagationVibrometryBased2025}.
Also, it has been applied to develop a model-updating workflow for thin-walled, damped duct structures, using Random Forest regressors as surrogate models in global optimization~\cite{heideggerModeorientedFiniteelementModel2026}.
Currently, pyCFS-data is used in the pre-processing of aeroacoustic source terms, e.g., of the aeroacoustic wave equation based on Pierce's operator \cite{AeroacousticSourcePotential,AcousticModelingUsing,schoderAeroacousticWaveEquation2022} and of the compressible perturbed convective wave equation~\cite{schoderCPCWEPerturbedConvective2023,schoderPerturbedConvectiveWave2026}. An aeroacoustic tutorial for beginners using OpenFOAM, pyCFS, and openCFS is provided by the test case FLOW-01~\cite{geyerFLOW01BenchmarkDataset2026}.

\section{Conclusions}
\label{sec:conclusions}
pyCFS-data~0.3.0 reads and writes openCFS HDF5 files into NumPy-based data structures, converts CFD, finite element, and measurement data from other formats, and provides a range of operators for mesh and data manipulations. Its restartable pipeline applies these operators to data sets larger than the main memory, which makes pyCFS-data a Python alternative to openCFS-data. This report summarized the structure of the software and the methods behind the operators to complement the user documentation.

\section*{CRediT authorship contribution statement}
\textbf{Andreas Wurzinger:} Conceptualization, Methodology, Software, Validation, Visualization, Writing -- original draft, Writing -- review \& editing.\\
\textbf{Patrick Heidegger:} Conceptualization, Methodology, Software, Validation, Visualization, Writing -- original draft, Writing -- review \& editing.\\
\textbf{Stefan Schoder:} Conceptualization, Funding acquisition, Project administration, Supervision, Writing -- review \& editing.

\section*{Declaration of generative AI and AI-assisted technologies}
During the preparation of this work Claude Code (Anthropic) with the model Claude Opus~5.5 was used to revise the manuscript for pyCFS~0.3.0, i.e., to draft and edit text and figures and to check the descriptions and the illustrative example against the source code. The tool was also used to assist the development of parts of the source code added in pyCFS~0.3.0.
After using this tool, the authors reviewed, tested, and edited the content and the code as needed and take full responsibility for the content of the publication.

\section*{Acknowledgments}
We thank all contributors of pyCFS and openCFS.

\section*{Funding information}
A.W. and S.~S. received funding from the COMET project ECHODA ("Energy Efficient Cooling and Heating of Domestic Appliances") under FFG project No. FO999913905. ECHODA is funded within the framework of "COMET - Competence Centers for Excellent Technologies" by BMIMI, BMAW, the province of Styria, and SFG. The COMET program is managed by FFG. S.~S. received funding from the FFG under project No. FO999913972. This research was funded in whole or in part by the Austrian Science Fund (FWF) 10.55776/PIN8859924. P.~H. and S.~S. received funding from the FFG project RETUNE under the project No. FO999937650.

\printbibliography

\end{document}